\documentclass[prl,preprint,aps,superscriptaddress,showpacs,floatfix]{revtex4}
\usepackage{amsmath,amssymb,latexsym,amsfonts}
\usepackage{epsfig}
\usepackage{graphics}
\usepackage{colordvi}

\newcommand{\nimb}{ Nucl.\ Instr.\ and Meth.\ Phys.\ Res.\ B }

\begin{document}
\title{Electron guiding through insulating nanocapillaries}
\date{\today}

\author{K.\ Schiessl}
\email[Electronic address: ]{klaus@concord.itp.tuwien.ac.at}
\affiliation{Institute for Theoretical Physics, Vienna University of Technology, Wiedner Hauptstra\ss e 8-10, A--1040 Vienna, Austria, EU}
\author{K.\ T\H{o}k\'{e}si}
\affiliation{Institute of Nuclear Research of the Hungarian Academy of Sciences, (ATOMKI), H--4001 Debrecen, P.O.Box 51, Hungary}
\author{B.\ Solleder}
\affiliation{Institute for Theoretical Physics, Vienna University of Technology, Wiedner Hauptstra\ss e 8-10, A--1040 Vienna, Austria, EU}
\author{C.\ Lemell}
\affiliation{Institute for Theoretical Physics, Vienna University of Technology, Wiedner Hauptstra\ss e 8-10, A--1040 Vienna, Austria, EU}
\author{J.\ Burgd\"orfer}
\affiliation{Institute for Theoretical Physics, Vienna University of Technology, Wiedner Hauptstra\ss e 8-10, A--1040 Vienna, Austria, EU}

\begin{abstract}
We simulate the electron transmission through insulating Mylar (PET) capillaries. We show that the mechanisms underlying the recently discovered electron guiding are fundamentally different from those for ion guiding. Quantum reflection and multiple near-forward scattering rather than the self-organized charge-up are key to the transmission along the capillary axis irrespective of the angle of incidence. We find surprisingly good agreement with recent data. Our simulation suggests that electron guiding should also be observable for metallic capillaries.
\end{abstract}

\maketitle

The field of charged-particle guiding through insulating nanocapillaries was initiated by the discovery of Stolterfoht \textit{et al.} \cite{hcicap_nico_prl02} that highly charged ions (HCI) could be transmitted through nanocapillaries formed in Mylar (PET) without undergoing charge exchange with the capillary walls. This was quite unexpected as transmission along the capillary axis could be observed for angles of incidence well outside the geometric opening angle $\theta_o$ given by the aspect ratio of the capillary (a typical aspect ratio of 1:100 corresponds to $\theta_o\sim 1^\circ$). This observation has been confirmed for other insulating materials \cite{saha,mat}. Apart from the conceptual interest into the underlying processes, guiding of ions also holds the promise to develop into an efficient tool to collimate and/or focus ion beams without the need for electrical feedthroughs with diverse applications, most notably to cell microsurgery \cite{iwai}. Guiding was interpreted in terms of a self-organized charge-up of the capillary wall \cite{hcicap_nico_prl02,hcicap_nico_pra07}. Microscopic simulations \cite{mycap_pra05,mycap_nimb05} revealed that after a distributed transient charge-up of the capillary wall a single or a few charge patches \cite{schuch} near the entrance dominate the guiding in dynamical equilibrium. The charging of the capillary wall acts as a Coulomb mirror which leads to elastic reflections from the wall (``trampoline'') at distances sufficiently large as to preclude charge transfer or electronic inelastic processes. One consequence of this scenario is that capillary transmission of keV HCI's proceeds not only in their initial charge state but also without any significant energy loss.

The very recent observation of a seemingly similar guiding effect for electrons through Al$_2$O$_3$ \cite{electronexp_serbia_pra06} and PET capillaries \cite{tanis} came as another surprise. Electrons are unlikely to encounter a Coulomb mirror as strong as in the case of HCI guiding. Secondary electron emission (SEE) coefficients for electron impact with a few hundred eV energy may suggest even positive charge-up resulting in attraction to rather than repulsion from the surface. Additionally, even in absence of any charge-up, the attractive long-range polarization potential (``image potential'') steers electrons towards the surface. This suggests that a fundamentally different guiding scenario must prevail. Indeed, first experimental data \cite{tanis} show a significant and, in many cases, dominant fraction of guided electrons having suffered considerable energy loss pointing to inelastic scattering events.

In this letter, we present the first microscopic simulation of electron transmission though insulating nanocapillaries within the framework of the mean-field classical transport theory (CTT) \cite{burg_ctt,deiss}. Within the CTT, it is possible to include quantum scattering effects via the collision kernel for the evolution of the ensemble of classical particles. One key ingredient to the understanding of electron guiding turns out to be quantum reflection at the attractive planar averaged surface potential (``planar channeling''). Another is the significant near-forward scattering probability for both elastic and inelastic scattering of electrons reaching and penetrating the internal wall of the capillary. In view of the complexity of the underlying processes, we find surprisingly good agreement with available data.

A full ab-initio simulation of the present multi-scale problem ranging from the atomic scale for electron-atom scattering ($\Delta x\lesssim 10^{-10}$ m) to the mesoscopic scale (length of the capillary $l\approx 10\, \mu$m) is clearly out of reach. We therefore perform the simulation within the framework of a mean-field classical transport theory \cite{burg_ctt} based on a microscopic classical-trajectory Monte Carlo simulation for electron transport. Accordingly, the classical Langevin equation
\begin{equation}
\dot{\vec{v}}=\vec F_{mean}(\vec r,t)+\vec F_{stoc}(\vec r,t)\label{eq1}
\end{equation}
is solved for a large ensemble of electron trajectories. The conservative force field $\vec F_{mean}$ accounts for the attractive image potential %($-\frac{1}{4d}\frac{\varepsilon -1}{\varepsilon+1}$)
near the surface as well as the charge-up of the capillary wall. The stochastic force $\vec F_{stoc}(\vec r,t)=\sum_i \Delta \vec p_i \cdot\delta(t-t_i)$ describes random momentum transfers $\Delta p_i$ at (almost) random times $t_i$. $\vec F_{stoc}$ can account for elastic and inelastic collisions electrons experience as they hit the internal wall of the capillary and penetrate into the surface layers of the bulk material. The energy $\Delta E$ lost in an inelastic scattering process is transferred to a secondary electron released at the position of the primary electron. Trajectories for these ``secondary'' electrons are followed as well and contribute to the total spectrum of electrons. The particle number for which the Langevin equation is solved is thus not conserved. The point to be stressed is that quantum properties of electron dynamics can be fed into the classical equation of motion via the collision kernel of the transport equation. Elastic scattering cross sections have been calculated with the ELSEPA code \cite{salvat} using bare atomic potentials for the insulator constituents. E.g.\ PET (sum formula C$_{10}$H$_8$O$_4$) is described as a compound containing about 45.5\% carbon and 18.2\% oxygen contributing to the elastic mean free path. The fraction of hydrogen atoms can be safely neglected due to the small elastic cross section of hydrogen as compared to the other constituents. Inelastic scattering is determined from the momentum and energy dependent dielectric function of the bulk material $\varepsilon_b (q,\omega)$ which is constructed from an extension of the optical data \cite{palik} for the capillary material [$\varepsilon_b(q=0,\omega)$] to the $q$--$\omega$ plane. Additionally, energy loss due to surface excitation has been included in our simulation. An approximate surface dielectric function $\varepsilon_s(q,\omega)$ is derived from the bulk dielectric function $\varepsilon_b(q,\omega)$ \cite{Ritchie,Reinh}.

\begin{figure}[h]
\epsfig{file=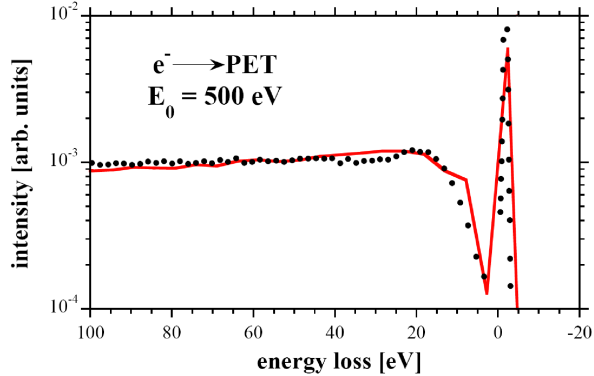,width=8cm,clip=}
\caption{(Color online) Energy spectrum for 500 eV electrons incident on PET under $\theta_{in}=40^{\circ}$ with respect to the surface. Solid circles show experimental data, solid lines results of our electron transport simulation. \label{fig2}}
\end{figure}

Electron impact on a flat PET surface provides a test for the reliability of the collision kernel employed in Eq.\ \ref{eq1}. Electrons with kinetic energies of 500 eV were directed on the target surface under an angle of incidence $\theta_{in}=40^\circ$ with respect to the surface. The measured spectrum of scattered electrons (Fig.\ \ref{fig2})
agrees remarkably well with our simulation thus lending creadence to our treatment of multiple scattering and collision cascades of penetrating trajectories.

The key novel feature for electron guiding is the glancing scattering at the planar-averaged surface potential $V_{pl}(z)$ of the capillary wall without penetrating into the bulk. This quantum reflection due to the attractive surface potential is completely absent in a truely classical simulation but can be included as a stochastic process into Eq.\ \ref{eq1}. The elastic specular reflection probability $P_s$ and momentum transfers $\Delta p_i=2k_\perp$ are determined for $V_{pl}(z)$ approximated alternatively by density functional theory (DFT) calculations of the target material and by a step function of the same height. For DFT calculations the program package ``ABINIT" \cite{abinit} is employed. We find that $P_s$ is generally larger for insulators than for metals as their surface potential is steeper (less electron spill-out). 
\begin{figure}[h]
\epsfig{file=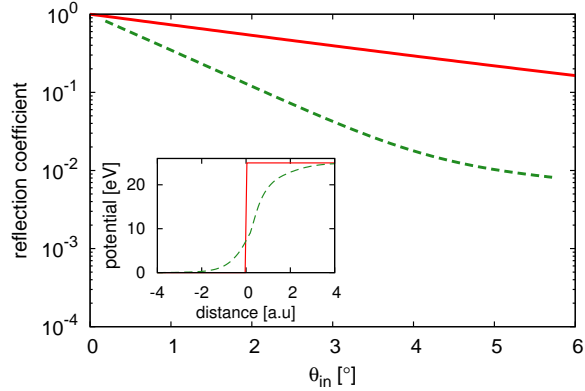,width=8cm,clip=} 
\caption{(Color online) 
Specular reflection coefficient $R_s$ as a function of $\theta_{in}$ for various types of surface potentials and a total kinetic electron energy of $E=0.5\cdot(k_\perp^2+k_\|^2)=500$ eV. Solid red line: step-potential, dashed green line: $V^{(DFT)}_{pl}(z)$ for PET \cite{pet_structure}. \label{fig1}}
\end{figure}
As an example, calculated values for $P_s$ are shown in Fig.\ \ref{fig1} as a function of $\theta_{in}$ (incidence angle with respect to the surface) for a kinetic energy of $E_{in}=500$ eV on a PET surface. A crystalline structure with a CH$_2$ group at the PET surface was assumed \cite{pet_structure}. To a good degree of approximation, $P_s(\theta_{in})$ can be fitted to an exponential, $P_s(\theta_{in})=\exp(-\theta_{in}/\theta_c)\label{eq3}$ with $\theta_c\approx 3.2^\circ$ for a step potential and $\theta_c\approx 0.95^\circ$ for a more realistic surface potential. In the limit $\theta_{in}\to 0$, $P_s$ converges to unity. 

\begin{figure}[h]
\epsfig{file=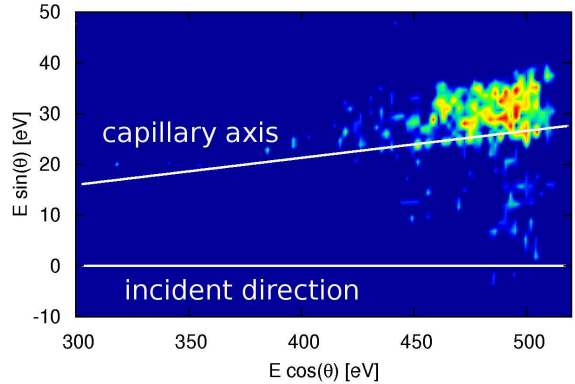,width=8cm,clip=}
\caption{(Color online) Distribution of guided electrons in the scattering plane incident with $E=500$ eV and angle $\theta_{in}=3^\circ$ relative to the capillary axis. As in \cite{tanis}, the incident beam has an energy spread of 20 eV full-width half maximum (FWHM), leading to a smooth cross-over from the elastic peak to the inelastically transmitted fraction. 
\label{fig3}}
\end{figure}

The two-dimensional distribution in the $E - \theta$ plane of 500 eV electrons incident at an angle of $\theta_{in}=3^\circ$ relative to the direction of the capillary axis for nanocapillaries with an aspect ratio 1:100 ($l=10\,\mu$m, diameter $D=200$ nm) unambiguously establishes guiding along the capillary axis of both elastically and inelastically transmitted electrons (Fig.\ \ref{fig3}).
Guided electrons that have suffered considerable energy loss (larger than the band gap) have penetrated the capillary wall and have undergone a multiple scattering sequence before reemerging in the open nanocapillary. Guiding is the result of the combined effects of the dominance of small-angle scattering for both elastic and inelastic scattering and of the drastically increased mean free path (or reduced extinction coefficient) when the scattered trajectory reenters the capillary. The peak of elastically guided elecctrons is of particular interest: while, unlike for HCI's, only a minor fraction of the total transmitted flux, it represents the closest analogue to ion guiding. In the present case, elastically guided electrons have undergone, on average, two to three collisions out of which at least one was an above-surface quantum reflection. While for inelastically guided electrons specular reflections are less prevalent, their contribution within the multiple scattering sequence is nevertheless important for determining the overall probability for guiding as well.

\begin{figure}
\epsfig{file=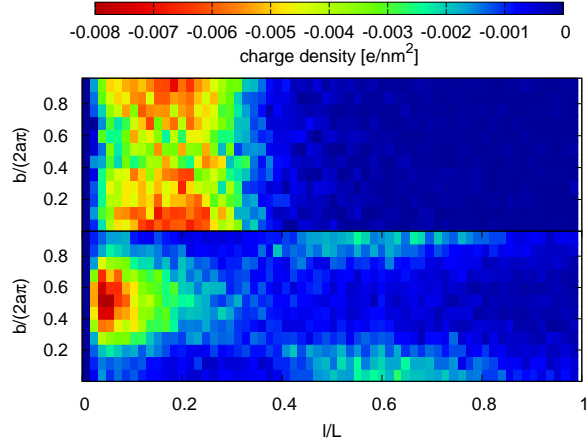,width=8cm,clip=}
\caption{(Color online) Equilibrium charge density $\rho$ on the internal walls of a PET capillary for $\theta_{in}=3^\circ$. The primary impact region of projectile electrons is situated at a cylinder arc length of $b=0.5/2\pi a$). The upper panel shows the result of the present simulation, the bottom panel shows $(-\rho)$ for ion guiding with charge $q=1$. The total charge accumulated on the capillary wall is in both cases $\sim 16000 |e|$.
\label{fig4}}
\end{figure}

To contrast the present novel scenario for guiding to that of ion guiding, it is of interest to inquire into the equilibrium charge-up of the capillary walls for both models (Fig.\ \ref{fig4}), i.e.,  we also performed a HCI simulation for ions with electron mass and charge $q=1$.
The major difference between the two models is the probability for depositing the projectile charge when hitting the capillary surface. While in the case of HCIs the charge is deposited at the impact point and spreads out due to surface and bulk transport, electrons can be scattered off the surface without leaving any charge behind or may even positively charge-up the impact area due to secondary electron emission. For small incidence angles (here $\theta_{in}=3^\circ$) the secondary electron coefficient is close to unity, $\delta \approx 1$, leaving the surface almost uncharged. Consequently, we find in the primary impact region a local minimum in the charge distribution for the present scenario (upper panel) as opposed to the maximum in the HCI simulation (lower panel). Moderate positive charge-up in the entrance area initially present in the electron simulation is greatly reduced by charge transport along the capillary walls in the equilibrium distribution. Secondary electrons with small energies ($E\lesssim 50$ eV) are recaptured at the opposite side of the capillary where they charge the surface negatively due to a small reflection coefficient and $\delta<1$. Impact on the surface following elastic and inelastic scattering events in the bulk leads to an almost uniform distribution over the downstream portion of the capillary surface with an almost even charge balance. An effective Coulomb mirror and, hence, electrostatic guiding is not operative. One immediate implication of this scenario is that no characteristic ``charge-up time'' is required before guiding sets in. On the contrary, the transmission rate is expected to be highest at $t=0$ and reduced with time due to charge-up (Fig.\ \ref{fig4}). Such a time dependence has, indeed, been observed \cite{milo} and is in clear contrast to the time dependence in the case of HCI transmission. Another consequence of the different charging characteristics is the energy spectrum of transmitted electrons: while energy loss is very small or completely absent if electrons are reflected at large distances from the surface (HCI model) direct interaction with the capillary wall leads to inelastic scattering events and substantial energy loss along the projectiles trajectory.

\begin{figure}
\epsfig{file=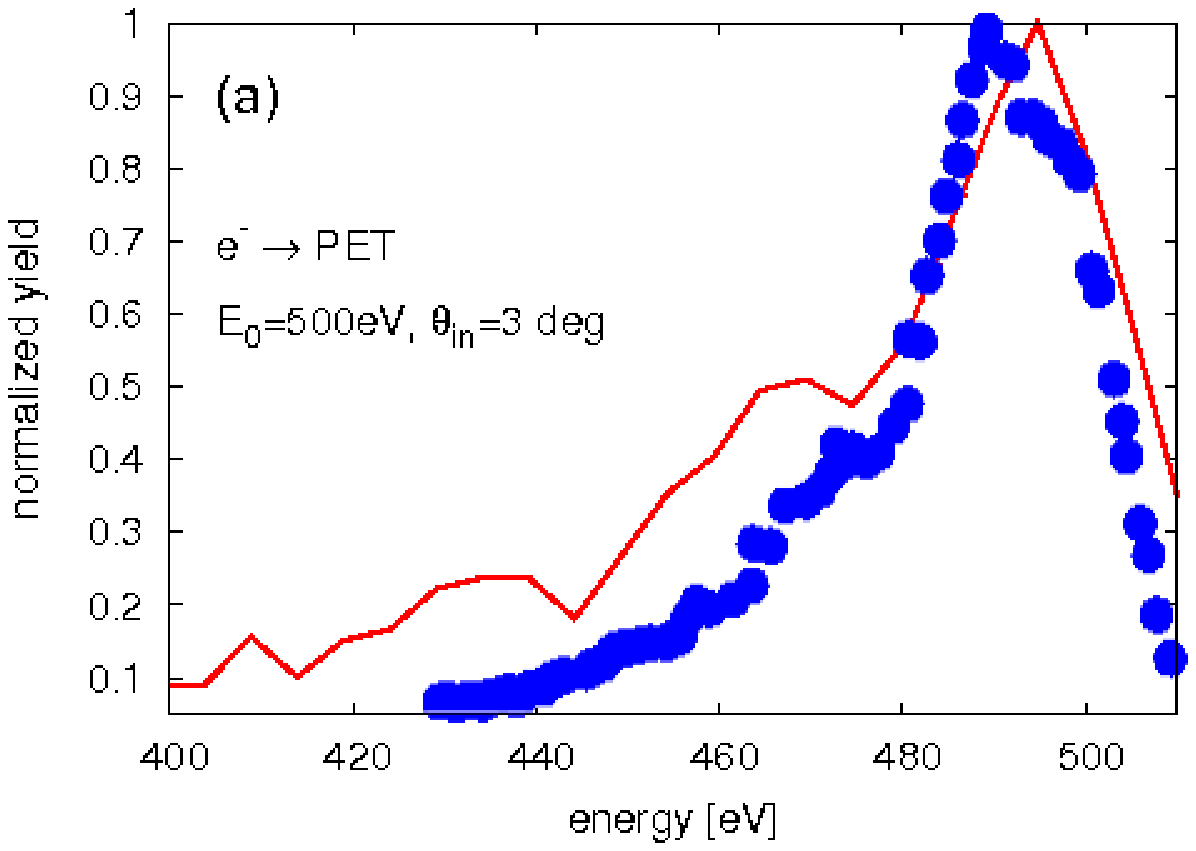,width=8cm,clip=}
\epsfig{file=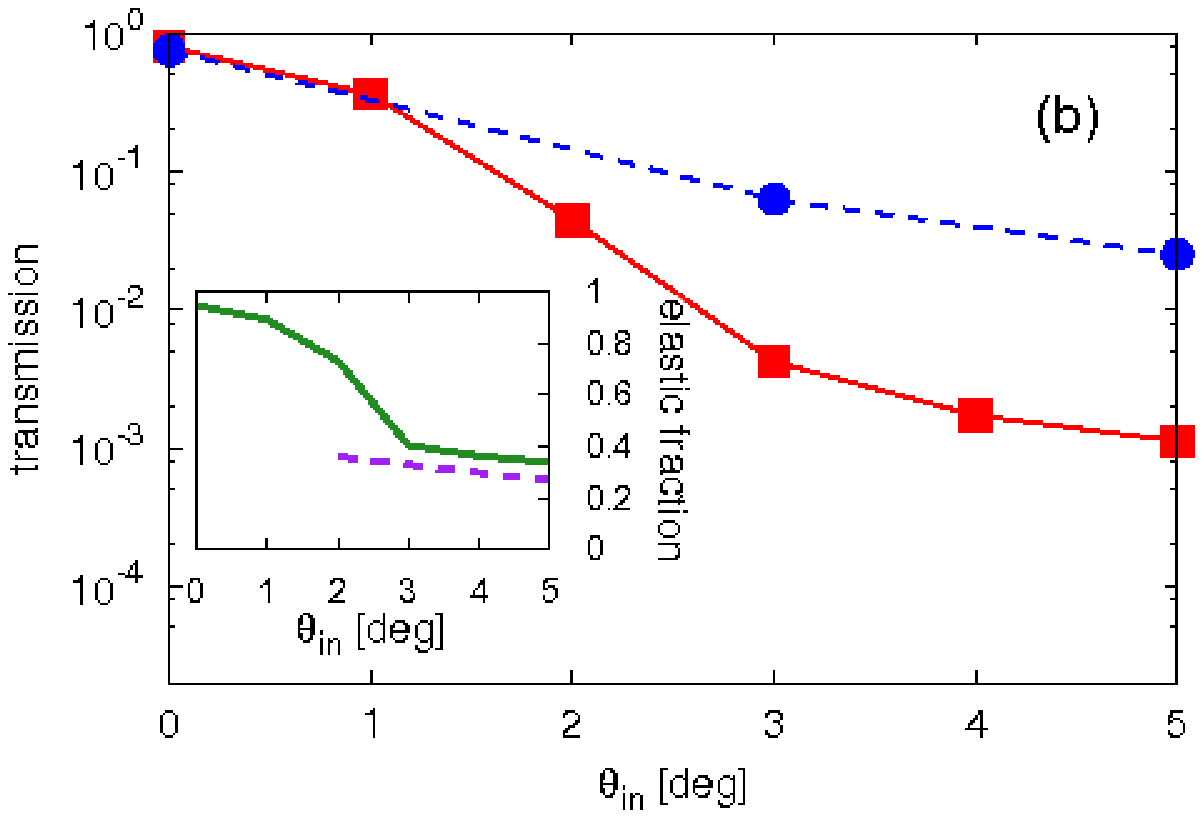,width=8cm,clip=}
\caption{(Color online) Energy distribution (a, linear scale) and transmission as a function of tilt angle (b, log scale) of guided electrons ($E_0=500$ eV, $\theta_{in}=3^\circ$). (a) solid circles: experimental data for an incident energy spread $\Delta E=20$ eV from \cite{tanis}; solid line: simulation. (b) Full symbols and dashed lines: experiment; solid lines: simulation. Transmission as a function of tilt angle irrespective of energy loss. The inset shows the fraction of (quasi-) elastically guided electrons. Experimental data (dashed line) is determined by subtracting the (estimated) inelastic component from the total yield (see \cite{tanis}).
\label{fig5}}
\end{figure}

A direct comparison with experimental data for PET yields, in view of the complexity of the process and the limited control over the geometry and surface composition of the internal capillary wall, remarkably and, possibly in part, fortuitously good agreement (Fig.\ \ref{fig5}).
A few discrepancies are worth noting. The lower energy tail corresponding to energy loss exceeding $\sim 20$ eV appears slightly overestimated (Fig.\ \ref{fig5}a). This may be due to the unknown surface composition of the target material resulting from the capillary-etching process. In our simulation a dielectric function for clean PET is used. The discrepancy in transmission at larger tilt angles (Fig.\ \ref{fig5}b) can, at least in part, be attributed to low-energy transmission ($E<20$ eV) which presently cannot be reliably modeled. These electrons may considerably contribute to the total number of electrons leaving the capillary. This is also evident in the fraction of electrons transmitted without energy loss. At larger tilt angles elastic fractions from simulation and experiment are in good agreement (inset in Fig.\ \ref{fig5}b, linear scale). At small tilt angles $\theta_{in}\leq 2^\circ$, direct transmission involving no collisions and specular reflection at the capillary wall are the dominant processes. Here, reliable experimental data for the elastic component are not yet available.

In summary, we have presented first microscopic simulations for electron guiding through nanocapillaries establishing a guiding scenario entirely different from that for highly charged ions. Quantal specular reflection at an attractive average surface potential and multiple small-angle elastic and inelastic scattering are key to guiding. Charge-up of the surface does play only a minor role in the guiding process as opposed to the case of highly charged ionic projectiles where strong electrostatic fields are required for guiding through insulating materials. One consequence of this scenario is the prediction that electron guiding should also be operational for other materials, in particular for metallic nanocapillaries. However, we expect reduced elastic transmission due to the smaller specular-reflection probability for metallic surfaces. We hope this hypothesis to be experimentally tested in the near future.
 
This work was supported by the Austrian \textit{Fonds zur F\"orderung der wissenschaftlichen Forschung} under grants no.\ FWF-SFB016 ``ADLIS'' and no.\ 17449 and the TeT Grant No. AT-7/2007. One of us (K.T.) was also partially supported by the grant ``Bolyai'' from the Hungarian Academy of Sciences and the Hungarian National Office for Research and Technology. K.S.\ acknowledges support by the IMPRS-APS program of the MPQ (Germany). The authors thank J.\ Toth from ATOMKI for providing us with experimental scattering data (Fig.\ \ref{fig2}).
%\end{acknowledgments}

\end{document}